\PassOptionsToPackage{dvipsnames}{xcolor}
\documentclass[manuscript,authorversion,nonacm]{acmart}

\AtBeginDocument{%
  }

\acmDOI{}

\usepackage{caption}
\usepackage{subcaption}
\usepackage{wrapfig}
\usepackage{expdlist}
\usepackage{times}   
\usepackage{flushend}
\newcommand{\bpstart}[1]{\vspace{1mm} \noindent{\textbf{#1.}}}
\newcommand{\quotes}[1]{``#1''}

\usepackage{siunitx}

\newcommand{\cm}[1]{\SI{#1}{\centi\meter}}

\newcommand{\revision}[1]{\textcolor{black}{#1}}

\begin{document}

\title{Visualization on Smart Wristbands:\\ Results from an In-situ Design Workshop with Four Scenarios}





\author{Alaul Islam}
\orcid{0000-0001-6900-3822}
\affiliation{%
  \institution{University Health Network}
  \city{Toronto}
    \state{Ontario}
  \country{Canada}}
\email{mohammad.alaulislam@uhn.ca}

\author{Fairouz Grioui}
\orcid{0009-0001-7358-6749}
\affiliation{%
  \institution{University of Stuttgart}
  \city{Stuttgart}
  \country{Germany}}
\email{fairouz.grioui@vis.uni-stuttgart.de}

\author{Raimund Dachselt}
\orcid{0000-0002-2176-876X}
\affiliation{%
  \institution{Technische Universität Dresden}
  \city{Dresden}
  \country{Germany}}
\email{raimund.dachselt@tu-dresden.de}

\author{Petra Isenberg}
\orcid{0000-0002-2948-6417}
\affiliation{%
  \institution{Université Paris-Saclay, CNRS, Inria}
  \city{Gif-sur-Yvette}
  \country{France}}
\email{petra.isenberg@inria.fr}

\renewcommand{\shortauthors}{Islam et al.}

\begin{abstract}
We present the results of an in-situ ideation workshop for designing data visualizations on smart wristbands that can show data around the entire wrist of a wearer. Wristbands pose interesting challenges because the visibility of different areas of the band depends on the wearer's arm posture. We focused on four usage scenarios that lead to different postures: office work, leisurely walks, cycling, and driving. As the technology for smart wristbands is not yet commercially available, we conducted a paper-based ideation exercise that showed how spatial layout and visualization design on smart wristbands may need to vary depending on the types of data items of interest and arm postures. Participants expressed a strong preference for responsive visualization designs that could adapt to the movement of wearers' arms. Supplemental material from the study is available here: \url{https://osf.io/4hrca/}.
\end{abstract}



\begin{teaserfigure}
  \includegraphics[width=1\linewidth]{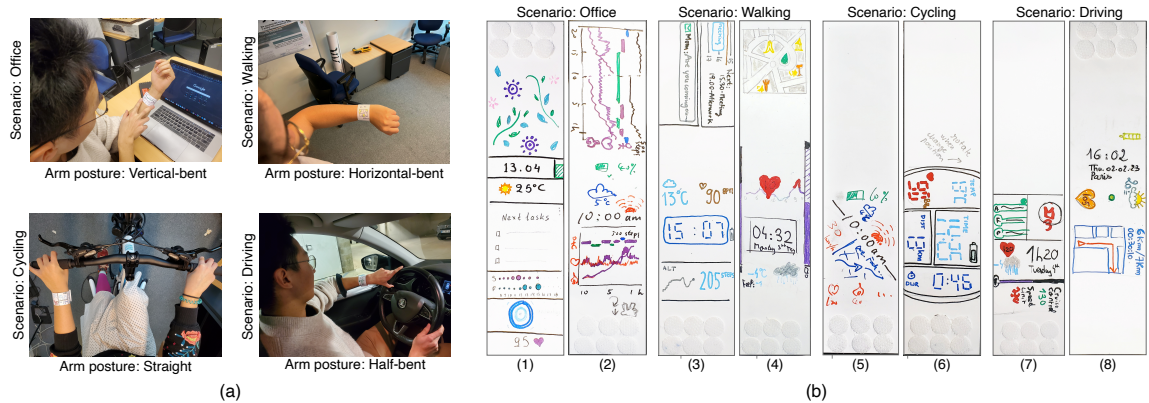}
  \caption{(a) Examples of the four arm postures taken by participants across four study scenarios: Vertical-bent (office), Horizontal-bent (walking), Straight (cycling), and Half-bent (driving). (b)
  Eight examples of wristband design sketches, two from each study scenario are shown. These examples showcase the use of various wristband zones; for example, sketches (1) and (2) in the office work scenario utilize all display zones, while sketches (3) and (4) in the walking scenario cover most of the display zones. Additionally, sketches (5) and (6) in the cycling scenario, along with (7) and (8) in the driving scenario, show the utilization of only a  few display zones. Participants expressed interest in responsive design, featuring dynamic changes based on arm posture (2), (5), and (6), the use of decorative elements (1), angled designs (5), (7), and (8), and detailed advanced visualizations (2).}
   \Description{The teaser of the paper. On the left side, it shows four pictures of the four arm postures taken by participants across the four study scenarios: (top-left) show a participant sitting in the office and looking at the wristband from a vertical-bent arm posture, (top-right) show a participant walking in a room while holding the wristband in a horizontal-bent arm posture, (bottom-left) shows a participant riding a bike and holding the bicycle handlebar with both hands creating a straight arm posture, and (bottom-right) shows a participant riding a car and looking at the wristband from a half-bent arm posture. On the right side of the teaser figure, we have eight examples of wristband design sketches, two from each study scenario are shown. These examples showcase the use of various wristband zones; from left to right, sketches (1) and (2) in the office work scenario utilize all display zones, sketches (3) and (4) in the walking scenario cover most of the display zones, sketches (5) and (6) in the cycling scenario, along with (7) and (8) in the driving scenario, show the utilization of only a few display zones.}
  \label{fig:teaser}
\end{teaserfigure}


\maketitle

\section{Introduction}

Curved smart wristbands share many characteristics with traditional wrist-worn devices such as smartwatches and fitness bands. Like these devices, they have displays worn around the wrist which can offer quick access to information. However, curved wristbands present significantly larger display areas---usually covering the wearer's whole wrist. Wristbands are interesting future technology that can overcome two important limitations of current smartwatches: the small display space and the fixed location of the display space on the wearer's arm. Traditional wrist-worn devices have a diagonal size of approx. 13-45\,\emph{mm}. Prior research demonstrated that this constrains the amount of displayed information---to 5 data items on average~\cite{Islam:2020:Smartwatch-Survey}, and indirectly influences the choice of data representations~\cite{Neshati_2019}, favoring simple representations such as text and icons over charts and graph visualizations. Additionally, the small size of the information displays negatively affects the reaction time of users \cite{Lyons_2016} in contrast to large smartband displays~\cite{Burstyn_2015}. Consequently, expanding the physical display would allow an increase in the number or size of the displayed representations and show labels, explanations, or currently unused chart types. 

In addition to increased display space, smart wristbands offer the benefit of allowing to display visualizations in locations tailored to the user's current context. For example, when riding a bike or holding a steering wheel, arms are constrained to postures that do not allow one to quickly glance at a smartwatch without rotating the wrist. Smart wristbands, however, could display information on the inside of the wrist to make it quickly glanceable. 

While some clear benefits of wristbands exist, we still lack formal study of the challenges and opportunities of designing visualizations for these displays, specifically taking different contexts of use---with their associated arm postures---into account. Therefore, we conducted an ideation workshop with 16 participants and asked them to sketch personalized designs of data visualization on a low-fidelity paper-based wristband in four distinct everyday life scenarios: sitting at an office desk, walking in a corridor, sitting on a bike, and sitting in a car. \revision{Our goal was to assess how different usage contexts can influence participants' data needs and how and where they prefer this data to be displayed on their wristbands. In addition, we recorded the varied arm postures participants adopted in different scenarios to ground our and future explorations of visualizations for smartbands in different scenarios.} Our analysis was guided by the following specific research questions:

\begin{description}[\compact\setlabelphantom{Q1:}\itshape]
\item[Q1:] What information would people primarily want to see on their wristbands? 
\item[Q2:] How do they envision this information to be presented? 
\item[Q3:] How and where do they envision visualizations to be placed on the wristband?
\item[Q4:] How do individual arm postures vary across different usage scenarios and how do postures influence envisioned data representation placement and design?
\end{description}

\section{Related Work}
\revision{\bpstart{Visualizations on Smartwatches} Smartwatches are increasingly used for personal data collection, and with the integration of Wi-Fi and Bluetooth connectivity, smartwatches have access to a wide range of data. The growing demand for display of this data drives research on the challenges of designing effective visualizations for small display sizes. Past work studied topics such as  the glanceability of smartwatch visualizations with low-level perceptual tasks~\cite{Tanja:2019:Glanceable-Visualization, blascheck:hal-04018448}, the influence of visual parameters like size, frequency, and color on reaction times~\cite{Lyons_2016}, and current representation types on smartwatch faces~\cite{Islam:2020:Smartwatch-Survey}. Additionally, some research has focused on designing visualizations for specific contexts, for example health~\cite{Amini:2017:Data-Representations-Health, Suciu:2018:Active-Self-Tracking-VAS} as well as design ideations~\cite{carpendale:2021:Mobile-Visualization-Design-Ideation, Gouveia:2016:Design-Space-Glanceable-Feedback, Cibrian:2020:Children-ADHD-Design-Challenges}. While there is evidence of a need for larger smartwatch form factors~\cite{Klamka2020} and studies showcasing the interaction benefits and effectiveness of larger wrist displays~\cite{Burstyn_2015}, fewer studies have explored visualization design for these expanded form factors.}

\bpstart{Large Wrist Displays} Past research explored various hardware designs of large wrist-worn displays with deformable and non-deformable materials. 
Some work used bendable displays that wrap around the whole or a big part of the wrist surface, for example electrophoretic \cite{Burstyn_2015}, e-paper \cite{epaper_2021, Strohmeier_2015}, and E-ink \cite{Snaplet_2011} displays. Others combined multiple small displays such as connected touch screens attached around the wrist \cite{Lyons_2012} creating a 360° display. The primary research emphasis of many of these works lay in validating the usability of the wrist display prototypes: when used for different functions~\cite{Snaplet_2011} (e.\,g., a worn watch or a flat handheld PDA) or under different interaction modalities (e.\,g., gesture-based~\cite{Burstyn_2015}, on display~\cite{Lyons_2012, Strohmeier_2015}, or beyond display~\cite{Saviot:2017:WRISTBAND_IO, Perrault:2013:Watchit}). Strohmeier et al.~\cite{Strohmeier_2015} found that users could quickly complete a task on a 3.5” large and cylindrical display spanning the entire wrist surface compared to a small one of 1.5”. The authors also emphasized that wearers could view a cylindrical display from any angle and were not limited to a single body pose, an important benefit for the representation of visualizations.  Saviot et al.~\cite{Saviot:2017:WRISTBAND_IO} and Perrault et al.~\cite{Perrault:2013:Watchit} expanded the interaction space of smartwatches to watch straps. However, discussions pertaining to the exhibited data and its representations often took a backseat, and the displayed information mainly served as a means to evaluate the display's performance within predefined usage contexts. In our study, our primary focus is the display of information on fully curved displays that are worn around the wrist. As no technology enabling quick design prototyping for these displays is yet available, we used paper-based bands resembling a sleeve-like curved display to conduct our ideation design workshop.

\bpstart{Displayed Data on Wristbands} 
The majority of data shown on wrist-worn curved display prototypes has been monochrome~\cite{Burstyn_2015, Lyons_2012, Strohmeier_2015, Snaplet_2011}. Klamka and Dachselt~\cite{epaper_2021}, however, used a bendable e-paper display with color capabilities, including six basic colors. The e-paper was used both as a stand-alone display and as a dynamic strap to extend a smartwatch display. The authors also tested eight usage scenarios with the e-paper prototypes including fitness activity tracking and showing weather data and notifications.  In another work, the same authors \cite{Klamka2020} explored different types of watchstraps with input and output capabilities (i.\,e., e-ink and a grayscale and a color OLED display) to visualize and interact with data. The authors conducted brainstorming and interview sessions for four applications (calendar, music, activity tracking, and navigation) to identify the different roles and data display distribution between the watch and the strap displays. The Watch+strap prototypes were designed to show various combinations of data types (e.\,g., fitness and navigation data) and representations (e.\,g., text, charts, and maps), in addition to different display layouts and interaction styles adapted to individual usage contexts. Similar to regular smartwatches, Facet~\cite{Lyons_2012} showed a set of different data (time, weather, email, and calendar entries) and representations (textual, illustration, and charts) spread over more than one screen worn as a bracelet. 

\revision{Previous research has not thoroughly investigated the wide range of wristband usage scenarios.} In our design workshop, we focus on four usage scenarios: office work, leisurely walks, cycling, and driving;  these contexts often have people moving or holding their arms in different postures. While prior work on wearables in these scenarios exists \cite{Burstyn_2015,Klamka2020,Lyons_2012} the impact on visualization choice, layout, and design has not been explored.

\section{Methodology}
\revision{It is important to anticipate solutions for future technologies to guide innovation, seize new opportunities, and outline potential challenges for adoption.} To elicit designs for technology that is not currently commercially available, dedicated ideation methodologies are needed. We chose to use an exploratory ideation methodology that emphasizes individual in-situ journaling~\cite{currier2018combining}. 
Our ideation exercise centered around sketching, a method that is described in~\cite{Greenberg2012Sketching} as \emph{a distinct form of drawing that supports the exploration and communication of ideas about designs}. Sketching allowed us to follow Kerzner and colleagues'~\cite {Kerzner2019Framework} recommendation to encourage participants to produce artifacts to externalize concepts. This approach facilitated subsequent documentation and idea analysis.


\bpstart{Study Design} 
Our ideation exercise asked participants to craft an extended watch face for a smart wristband. A watch face is a continuously visible interface from which users can glean essential information. Participants sketched four wristband designs, one for each usage scenario (\autoref{fig:teaser}(a)\,). The goal was to ascertain whether diverse activities could inspire a spectrum of wristband face designs tailored to each context. All participants took part in the four usage scenarios. To mitigate fatigue and order bias, we counterbalanced the sequence of scenarios. The workshop was conducted in our university lab at an office desk, in the hallway for walking, at a stationary bike, and in a parked car. Each design session lasted 60 minutes on average.

\bpstart{Participants}
We recruited 16 participants: 10 females and 6 males. Their age ranged between 24 and 34 years. All participants were from our university, including 5 Master's students, 8 PhD students, and 3 researchers. \revision{Except for one participant, all were familiar with the field of human-computer interaction, and eight participants had prior experience in designing data visualizations. We began the workshop by explaining what a smart wristband is and how it works. Participants were not required to own or wear a smartwatch to participate in the workshop.}


\bpstart{Procedure}
Before starting the design process, we fit a \cm{4}-wide white construction paper around participants' wrists (see \autoref{fig:teaser}(b)\,). For each of the four scenarios participants spent the first 2--3 minutes envisioning themselves engaged in the activity that they were going to design for (e.\,g., working at the office desk or holding the handlebars of a bike) while wearing the wristband prop. Then, they showed us the different arm postures they would adopt in the scenario if they were to view potentially displayed data. 
Next, we asked them to sketch an extended watch face on the paper prop they were wearing featuring time (analog or digital), smart wristband battery level, temperature, and heart rate; which is data commonly shown on smartwatches \cite{Islam:2020:Smartwatch-Survey}. In addition, participants chose at least two additional data types to sketch. 
After completing their initial sketches, participants removed their wristbands and spent 4 minutes refining their drawings.
Upon completion, one author of the paper reviewed designs with participants and took notes on their comments. 
Finally, participants wore the wristband again to \revision{demonstrate} their designs aligned with their intentions, i.\,e., while riding a stationary bike, sitting at an office, sitting in the driver's seat without actually driving, and walking along a corridor. By the end of each step, we captured photos of participants' arms and their designs. \revision{In this phase, participants did no longer adapt their designs.} 

\section{Findings}

 To analyze the 64 sketches, \revision{three of the authors} conducted an open card sorting exercise~\cite{Wood2008CardSorting} with printed copies of the sketches. The dimension of each image was 25 cm × 11 cm. We annotated different aspects of the design sketches for each participant and condition, including preferred data items, data representation types, chosen display zones for data representation, viewing arm posture, any mentioned rotation angle and type, and any comments they made. Here, we present our findings.

\subsection{Q1: What information did people primarily want to see on their wristbands?}

\begin{figure}[tb]
    \centering
    \includegraphics[width=.5\linewidth]{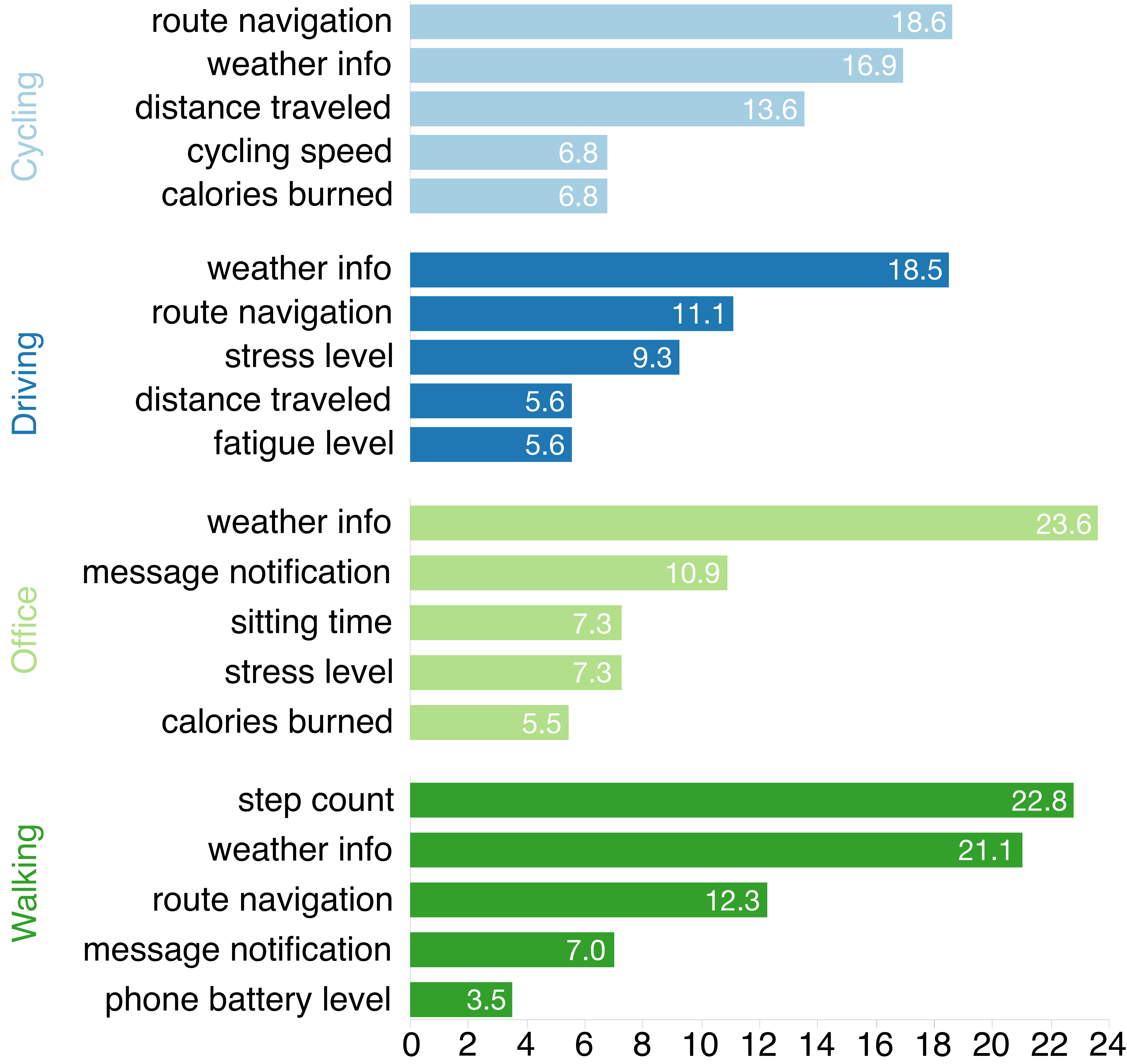}
    \caption{The distribution (in \%) of the top five data items in four study scenarios: office, walking, cycling, and driving reported by the participants.}
     \Description{The figure shows a horizontally grouped bar chart. The graph depicts the distribution (in \%) of the top five data items over the four study scenarios from top to bottom: cycling, driving, office, and walking as reported by the participants. For cycling, route navigation is the most reported data item, with 18.6\%, followed by weather info (16.9\%) and distance traveled (13.6\%). For driving and office scenarios, weather info is first ranked with 18.5\% and 23.6\%, respectively. Route navigation and stress level are second (11.1\%) and third (9.3\%) for the driving scenario, and message notification and sitting time are second (10.9\%) and third (7.3\%) for the office scenario. Finally, for the walking scenario, step count was the most reported data item, with 22.8\%, followed by weather info (21.1\%) and route navigation (12.3\%).}
    \label{fig:top_five_data_items_by_scenarios}
\end{figure}

\begin{figure}[tb]
    \centering
    \includegraphics[width=.6\linewidth]{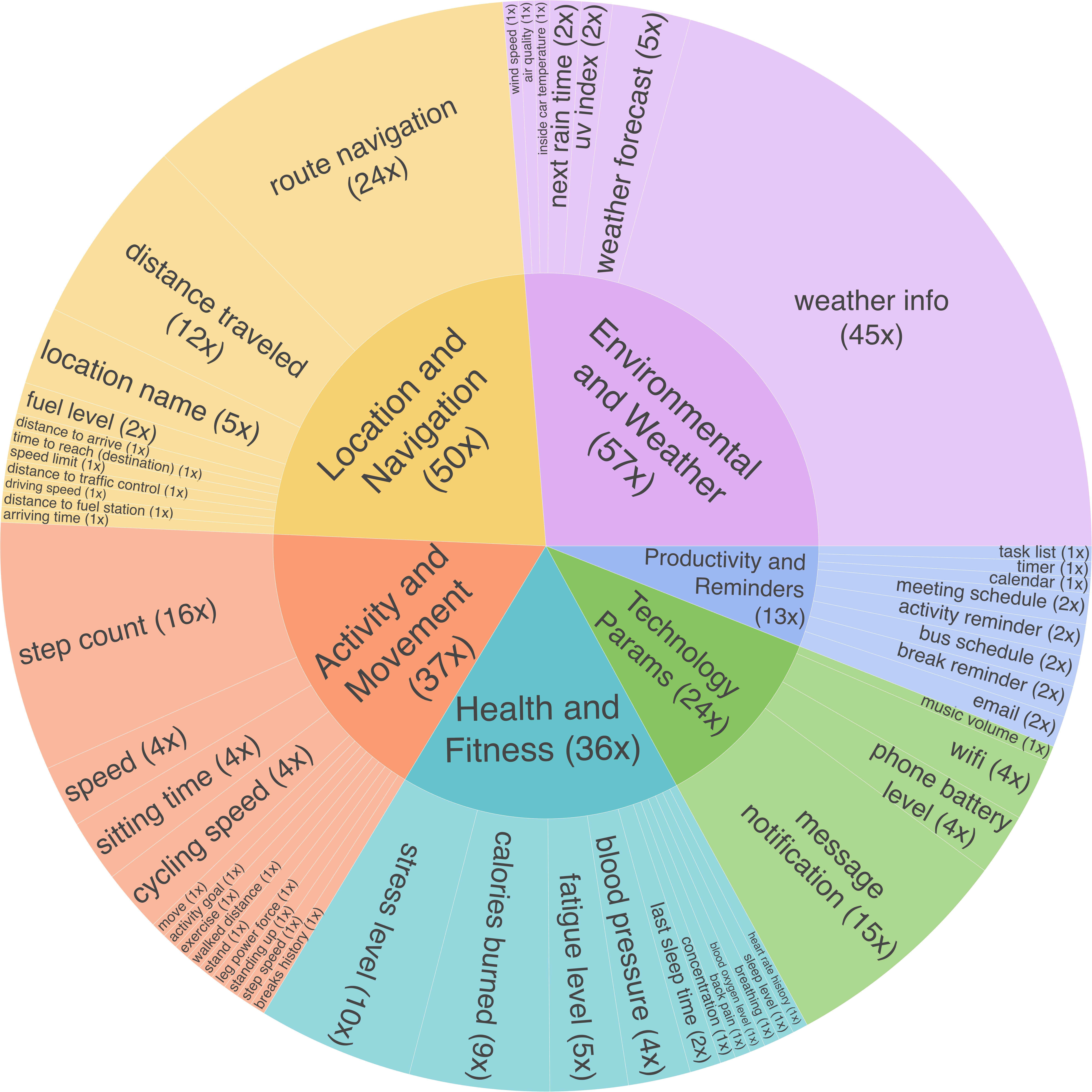}
    \caption{The participants reported data items according to six data categories in addition to the mandatory four data items (time, temperature, heart rate, and watch battery level).}
    \Description{The figure shows a sunburst chart depicting all the data items reported by participants. The data items are categorized according to six categories, including environmental and weather data (57 data items), location and navigation (50 data items), activity and movement (37 data items), health and fitness (36 data items), technology and params (15 data items), and productivity and reminders (13 data items).}
    \label{fig:all_data_items_sunburst}
\end{figure}

Apart from the four mandatory data items (time, battery level, temperature, and heart rate) that we instructed participants to include, they were also required to select additional data items based on their preferences. This additional data varied depending on the context of use (see \autoref{fig:top_five_data_items_by_scenarios}). For instance, route navigation data appeared as a top choice in activities involving movement: 18.6\% of the time in the cycling context, 12.3\% in walking, and 11.1\% in driving. Other data was chosen in all scenarios: weather information on sky conditions consistently ranked as one of the most frequently desired data items, holding the first or second position across all activities. Similarly, health and fitness data, including step count, calories burned, and stress level, were commonly chosen in all scenarios.
To uncover higher-level insights and patterns on the data, we aggregated all additionally chosen data items into six distinct groups (see \autoref{fig:all_data_items_sunburst}). Environmental \& Weather data emerged as the most common category, chosen 57\texttimes, followed by Location \& Navigation (50\texttimes), Activity \& Movement (37\texttimes), Health \& Fitness (36\texttimes), Technology Parameters (24\texttimes), and Productivity \& Reminders (13\texttimes).  Interestingly, we found that health and fitness-related data was frequently reported in office scenarios, hinting that individuals are indeed mindful of their health and fitness while working in the office.

\subsection{Q2: How was the information represented?}

\begin{figure*}[t!]
\includegraphics[width=.4\textwidth]{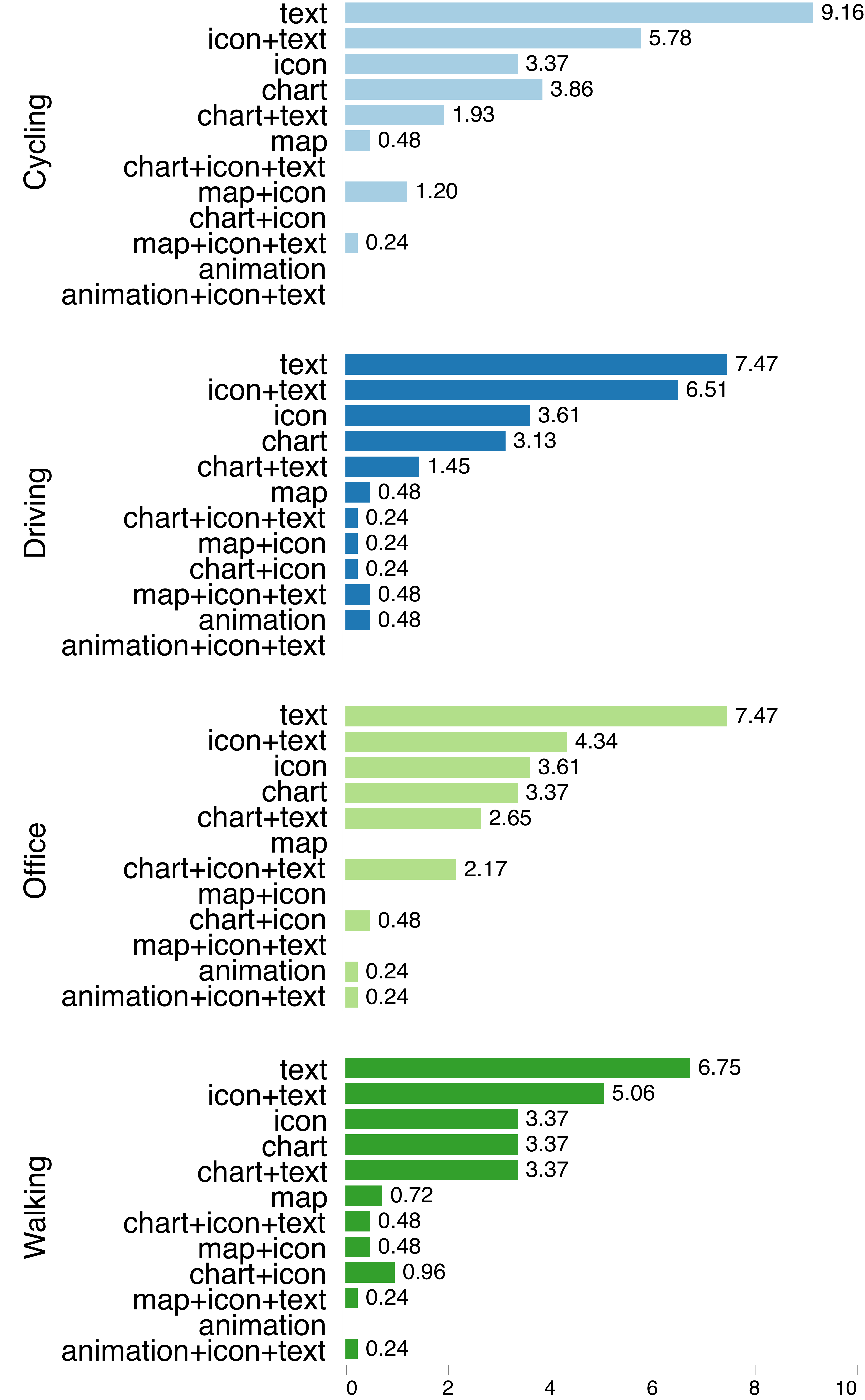}
  \includegraphics[width=.4\textwidth]{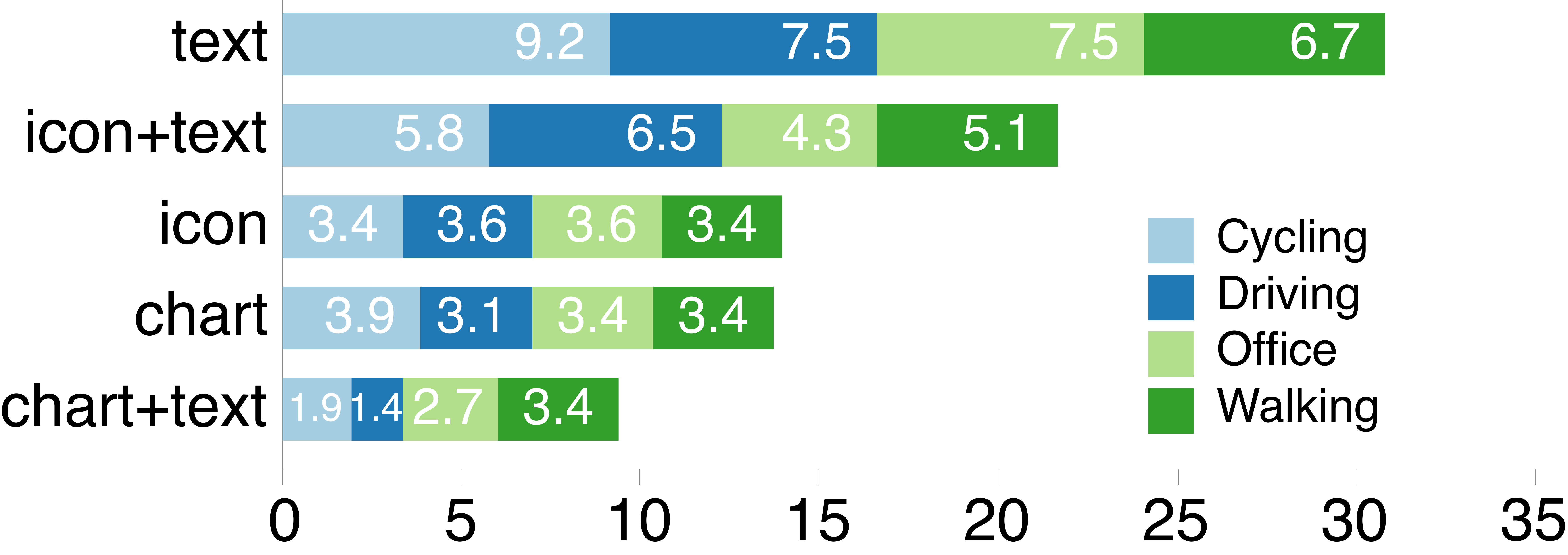}
  \caption{(Left) The distribution (in \%) of 12 data representation types in four study scenarios: office, walking, cycling, and driving reported by the participants. (Right)
  The distribution (in \%) of the top five data representation types from 12 types reported in four scenarios.}
 \Description{The figure shows two graphs: On the left side, a horizontally grouped bar chart that depicts the distribution (in \%) of 12 types of data representation over the four study scenarios from top to bottom: cycling, driving, office, and walking. For all scenarios, the top two data representation types are text and icon+text, followed by chart, then icon for cycling, and icon, then chart for driving and office scenarios. For the walking scenario, all icon, chart, and chart+text representations are ranked third with equal percentage of 3.37\%. The graph on the right side is a horizontal stacked bar chart. It shows the distribution (in \%) of the top five data representation types, from top to bottom: text, icon+text, icon, chart, and chart+text over the four scenarios. It shows that text and chart representation were mostly used during the cycling scenario with 9.2\% and 3.9\%, respectively. Also, the icon+text was mostly used in the driving (6.5 \%), and icon representations were equally used in driving and office scenarios with 3.6\%. Finally, the chart+text representation was the most used in the walking scenario (3.4\%).}
   \label{fig:data_representation}
\end{figure*}

\bpstart{Representation types} Our analysis showed that participants used a total of 12 different data representations across our four study scenarios (see \autoref{fig:data_representation}, (left)\,). Considering the top five data representation types (see \autoref{fig:data_representation}, (right)\,), text appeared as the primary way of data representation across all scenarios with a 30.9\% of the total representation types.  
Also, the use of icons was very frequent, either shown alone (14\%) or in combination with other forms of representation, such as maps, charts, and animations. While representations only consisting of charts were less common (13.8\%), they were often combined with other forms, including text (9.4\%), icons (1.7\%), or both (2.9\%), in four scenarios. Our results show that map-based representations, alone or combined with icons or text, were exclusively reported to show navigation itineraries for scenarios that required long-distance traveling. Finally, participants' designs rarely envisioned animations. When they were used, they appeared primarily in the driving and office scenarios. Participants explicitly mentioned using animation by combining written descriptions with accompanying designs.


\bpstart{Rotation of the representation}
Looking at participants' designs, it was clear that many of them counted on the responsiveness of the display to make the information reading convenient by adjusting the visualization rotation angles. 


Participants rotated the representations in two main angles: 90° and 45° from the initial viewing posture they defined. In general, the 90° rotation was used for 27 designs (9 for cycling, 8 for driving, 5 for walking, and 5 for the office) compared to 4 designs with the 45° rotation angle (2 for cycling, 1 for driving and 1 for walking). We also noticed that while the majority (23 designs) of the rotations described by participants were static, i.\,e., the display rotation was fixed and independent of the arm movements, only 8 designs were dynamically rotated, i.\,e., the display rotates following the wearer's arm movements in real time.

\subsection{Q3: How were the representations spatially laid out?}

\begin{figure}[tb]
    \centering
    \includegraphics[width=1\linewidth]{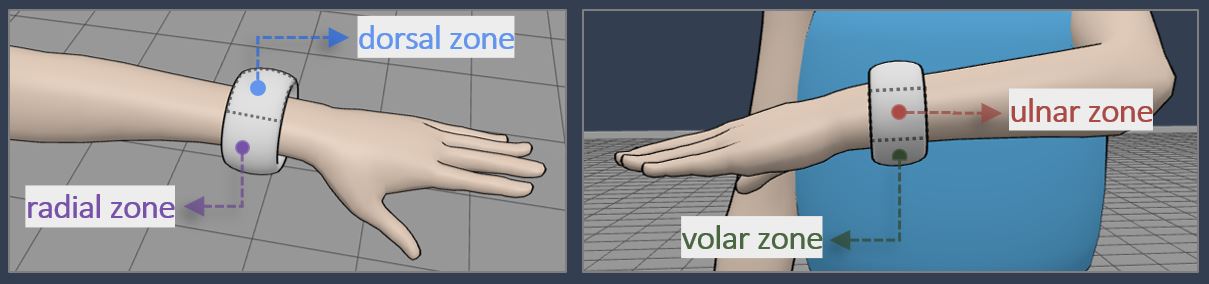}
    \caption{\small Illustrations showing the four display zones: {\textcolor{CornflowerBlue}{dorsal}, \textcolor{ForestGreen}{volar}, \textcolor{RoyalPurple}{radial}, and \textcolor{BrickRed}{ulnar} zones}, resulting from our card sorting analysis.}
     \Description{The figure shows two Illustrations of 3D models of a person wearing the wristband. The wristband is captured from different perspectives to show the four display zones: dorsal (top), volar (bottom), radial (right-side), and ulnar (left-side) zones.}
    \label{fig:display-zones-designs}
\end{figure}

\begin{figure*}[t!]
\centering
\includegraphics[height=.5\textwidth]{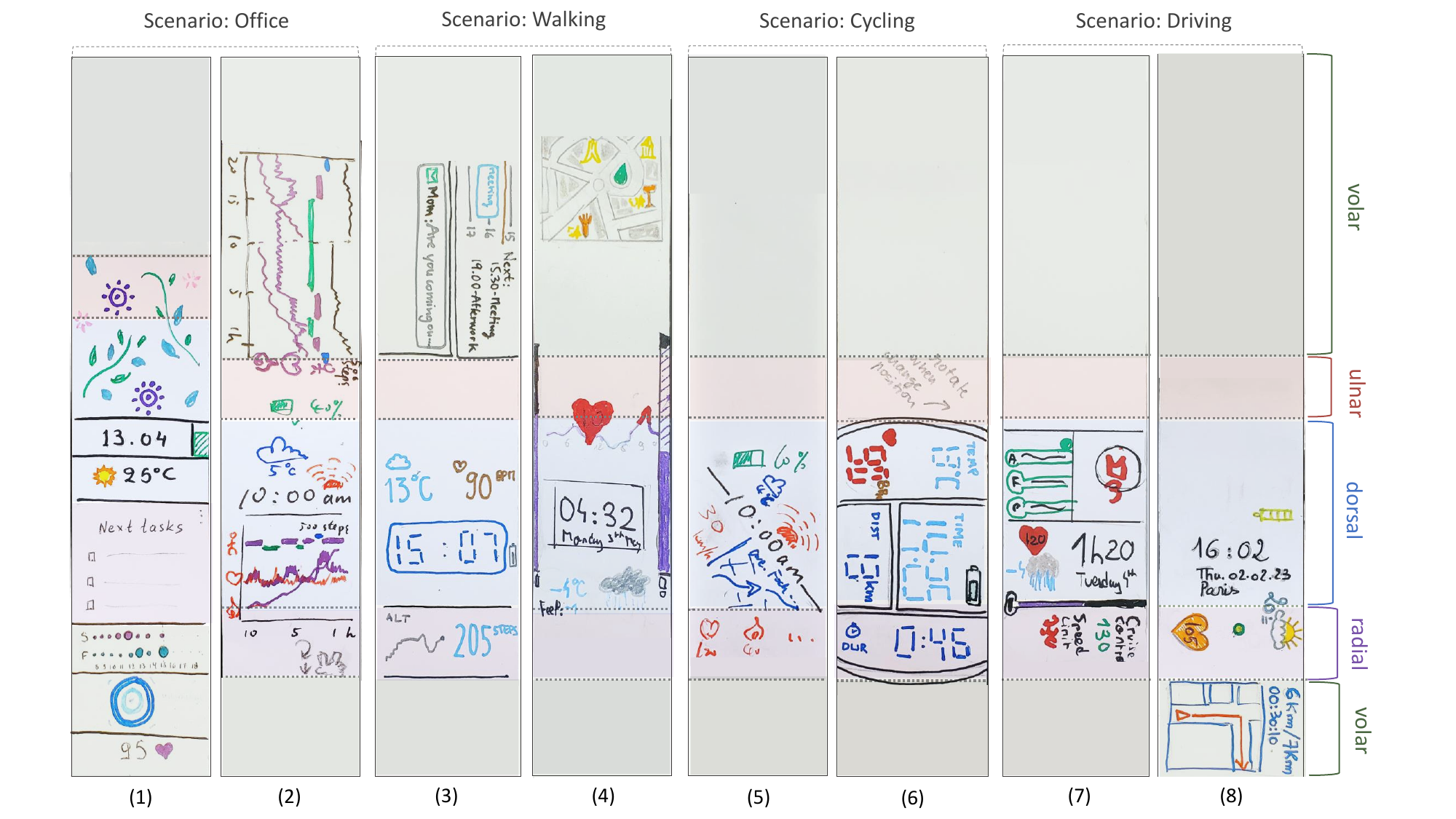}
  \caption{Examples of eight sketches designed by our participants for the four scenarios. For each sketch we highlighted the four display zones: volar, ulnar, dorsal, and radial to delimit where the data was displayed and viewed by participants.
  }
   \Description{The figure shows eight different examples of sketches designed by our participants for the four scenarios. For each sketch, we highlight the four display zones: volar, ulnar, dorsal, and radial to delimit where the data is displayed and viewed by participants.}
  \label{fig:sketches_display_zones}
\end{figure*}

\begin{figure}[tb]
    \centering
    \includegraphics[width=.5\linewidth]{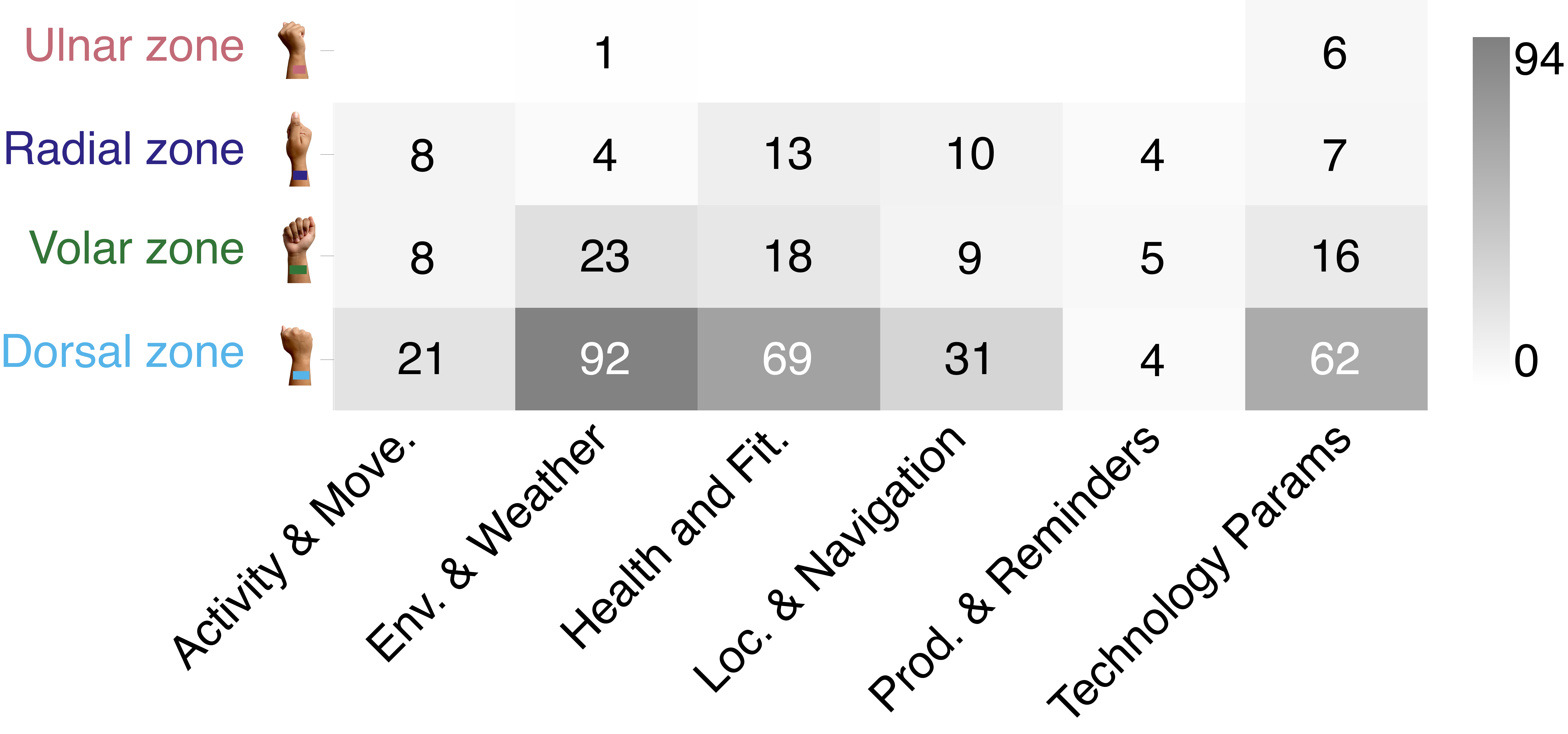}
    \caption{Spatial layout preferences reported by display zones: \textcolor{BrickRed}{ulnar}, \textcolor{RoyalPurple}{radial}, \textcolor{ForestGreen}{volar}, \textcolor{CornflowerBlue}{dorsal} and by data categories: Activity \& Movement, Environmental \& Weather, Health \& Fitness, Location \& Navigation, Productivity \& Reminders, and Technological Parameters.}
    \Description{The figure shows a heatmap depicting spatial layout preferences reported by display zones: ulnar, radial, volar, and dorsal, and by data categories: Activity \& Movement, Environmental \& Weather, Health \& Fitness, Location \& Navigation, Productivity \& Reminders, and Technological Parameters. The graph shows that the ulnar zone is the least used display zone (environment and weather:1x and technology and params:6x). Also, most data items are displayed on the dorsal zone (activity and movement:21x, environment and weather:92x, health and fitness:69x, location and navigation:31x, technology and params:62x). The only exception is with the productivity and reminder data items that are shown on the volar zone more frequently (5x), with a very slight difference to other zones (4x).}
    \label{fig:display-zones-data}
\end{figure}

Our analysis showed that participants placed the representations in four primary locations: the \textcolor{CornflowerBlue}{dorsal} zone, \textcolor{ForestGreen}{volar} zone, \textcolor{RoyalPurple}{radial} zone, and \textcolor{BrickRed}{ulnar} zone as shown in \autoref{fig:display-zones-designs}. The majority of data was displayed in the \textcolor{CornflowerBlue}{dorsal} zone (279\texttimes)---which is where people typically wear a watch display---followed by the \textcolor{ForestGreen}{volar} zone (79\texttimes) (below the wrist), \textcolor{RoyalPurple}{radial} zone (46\texttimes) (inside, thumb side), and \textcolor{BrickRed}{ulnar} zone (7\texttimes) (outside) for all types of data categories (\autoref{fig:display-zones-data}). \revision{We found no substantial influences of four mandatory data items (time, battery level, temperature, and heart rate) on representation designs sketched by the participants except zone preference for displaying the mandatory data items. These mandatory data items appeared more frequently in the \textcolor{CornflowerBlue}{dorsal} zone than in other zones.}
The \textcolor{CornflowerBlue}{dorsal} zone was mostly used to display Environmental and Weather data items (e.\,g., weather forecast and weather info), followed by Health and Fitness data (e.\,g., stress level, burned calories, and blood pressure), and Technological parameters (e.\,g., WiFi, phone battery level, and notifications). However, Productivity and Reminder data (e.\,g., email and activity reminder) were less common in this zone. \autoref{fig:display-zones-data} shows similar trends for the \textcolor{ForestGreen}{volar} zone but with approx. 4\,\texttimes fewer data items compared to the \textcolor{CornflowerBlue}{dorsal} zone. In general, the \textcolor{CornflowerBlue}{dorsal} zone was most favored for summarized data (e.g., current heart rate values (see \autoref{fig:sketches_display_zones}, (4, 6, 7)\,). At the same time, the \textcolor{ForestGreen}{volar} zone showed more detailed data summaries (e.g., heart rate value charts throughout the day such in \autoref{fig:sketches_display_zones}, (2)\, or map navigation as shown in \autoref{fig:sketches_display_zones}, (4) and (8)\,). The \textcolor{RoyalPurple}{radial} zone was considered next in priority for data placement, which varied depending on the specific tasks participants were engaged in, with a focus on Health and Fitness data. In contrast to the two previous zones, the \textcolor{RoyalPurple}{radial} zone rarely showed Environmental and Weather data. Finally, the \textcolor{BrickRed}{ulnar} zone received minimal attention from participants and was mainly limited to display technological parameters. In a few cases, the visualizations were displayed across display zones. For instance, \autoref{fig:sketches_display_zones}, (1) shows that the decoration displayed at the top of the \textcolor{CornflowerBlue}{dorsal} zone continues to the \textcolor{BrickRed}{ulnar} zone. Likewise, \autoref{fig:sketches_display_zones}, (2) and (4) show examples of sketches where part of the visualizations on the \textcolor{CornflowerBlue}{dorsal} zone are partially viewed on the \textcolor{BrickRed}{ulnar} and slightly intersect the \textcolor{RoyalPurple}{radial} zones.

\subsection{Q4: How did individual arm postures vary across different usage scenarios?}

\begin{figure}[tb]
    \centering
    \includegraphics[width=.5\linewidth]{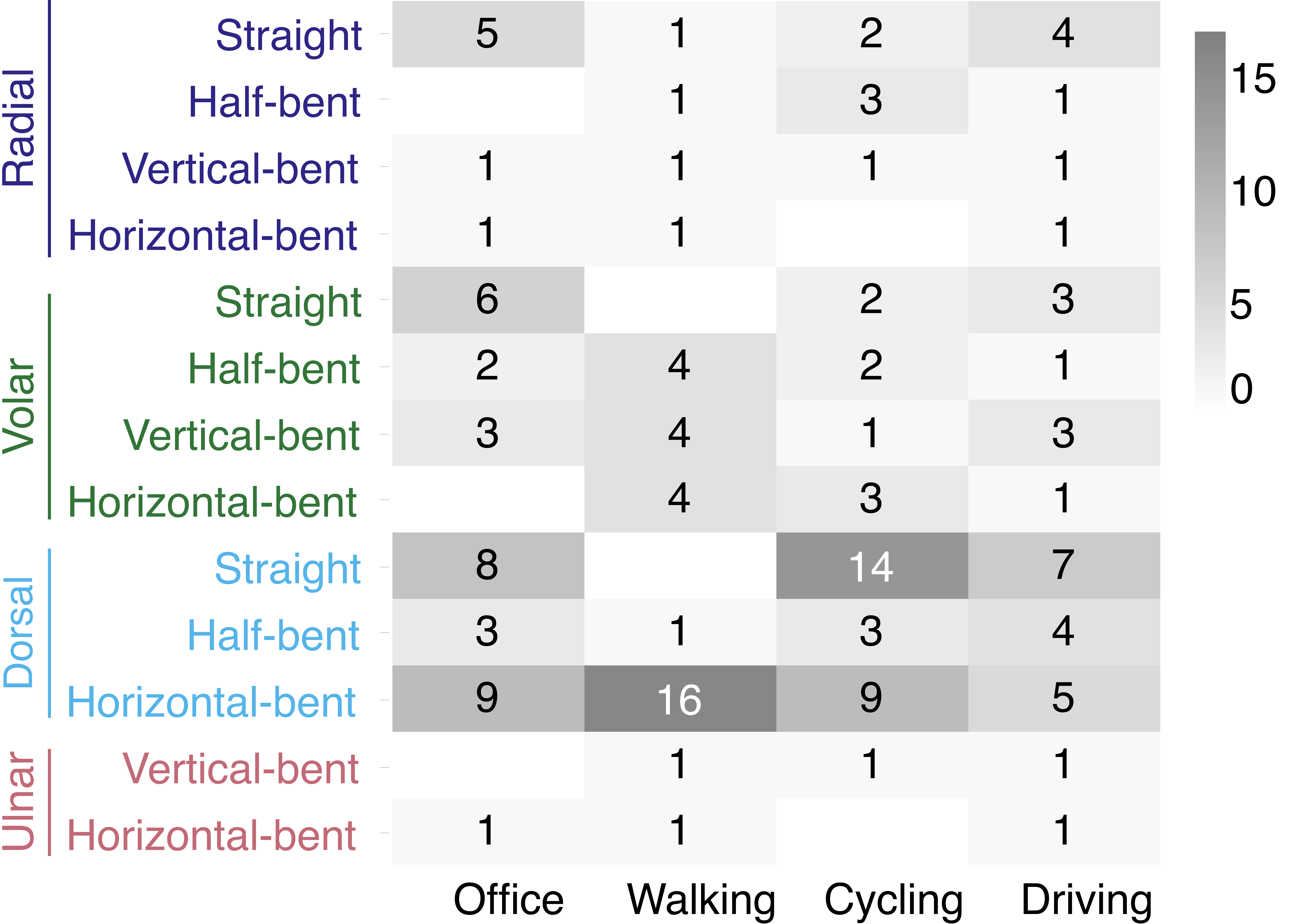}
    \caption{Count of arm postures performed by participants across four spatial display zones: \textcolor{BrickRed}{ulnar}, \textcolor{CornflowerBlue}{dorsal}, \textcolor{ForestGreen}{volar}, and \textcolor{RoyalPurple}{radial} zones, and arm postures: straight, half-bent, horizontal-bent, and vertical-bent, in four study scenarios: office, walking, cycling, and driving.}
    \Description{The figure shows a heatmap of the count of arm postures: straight, half-bent, horizontal-bent, and vertical-bent, performed by participants across the four spatial display zones: ulnar, dorsal, volar, and radial zone and over the four study scenarios: office, walking, cycling, and driving. The graph shows that in the dorsal zone, the horizontal-bent posture was prevalent during walking scenarios (16×), including the highest frequency overall (9+16+9+5 = 39\texttimes) across all scenarios. The second most common posture was straight (8 + 14 + 7 = 29\texttimes) in the dorsal display zone, with its highest occurrence observed during cycling scenarios (14×) and none during walking scenarios. Participants did not adopt the vertical-bent arm posture to read information in the dorsal zones. Additionally, in the ulnar zone, only two arm postures were used by a few participants across all usage scenarios: vertical-bent and horizontal-bent.}
    \label{fig:display-zones-posture}
\end{figure}

We categorized participants' arm postures into four main configurations including: \emph{straight, half-bent, horizontal-bent,} and \emph{vertical-bent}~(\autoref{fig:teaser}, (a)\,). We analyzed pre-sketch and after-sketch arm postures and compared posture types to the zones in which particicipants envisions their designs to appear. Unsurprisingly, participants' arm postures varied depending on the scenarios. Moreover, in many cases, multiple arm postures were used to view the information displayed on the same zone~(\autoref{fig:display-zones-posture}). For example, when analyzing data in the \textcolor{CornflowerBlue}{dorsal} zones, the horizontal-bent posture was prevalent during walking scenarios (16\texttimes), including the highest frequency overall ($9+16+9+5=39$\texttimes) across all scenarios. The second most common posture was straight ($8+14+7=29$\texttimes), which led to more information being put into the \textcolor{CornflowerBlue}{dorsal} display zone, with its highest occurrence observed during cycling scenarios (14\texttimes) and none during walking scenarios. The participants used all four arm postures while displaying information in the \textcolor{RoyalPurple}{radial} and \textcolor{ForestGreen}{volar} layouts. Participants did not adopt the vertical-bent arm posture to read information in the \textcolor{CornflowerBlue}{dorsal} zones. To read information in the \textcolor{BrickRed}{ulnar} zone, we observed only two arm postures by a few participants across all usage scenarios: vertical-bent and horizontal-bent.

\section{Discussion and Future Work}
In this section, we summarize the lessons learned from our ideation workshop into concrete design considerations.

\bpstart{Smartwatch research can serve as inspiration and comparison} 
The data items that participants chose for their design and their representations were similar to those used on regular smartwatches and fitness bands \cite{Islam:2020:Smartwatch-Survey,islam:2024:DesignSpace}; with some notable exceptions. Weather-related data in the smartband designs was more common than previously found on smartwatches where health and fitness-related data dominated \cite{Islam:2020:Smartwatch-Survey,islam:2024:DesignSpace}. We explain this difference by the fact that sky conditions were often (45\texttimes) put next to temperature data, which participants were required to display. In addition, some data on smartwatch faces appeared often on the bands related to the usage context: route navigation (for cycling, driving, walking), stress level (driving, in the office), or fatigue level (driving). Yet, overall, participants cared about many similar types of data as smartwatch wearers, including distance traveled, step counts, or notifications.

\bpstart{Create context-dependent display zones}
It was obvious in our data analysis that participants had created specific zones to lay out their data and that visualizations rarely crossed the boundaries of these zones. Many preferred the  \textcolor{CornflowerBlue}{dorsal} zone of the wrist, likely due to familiarity with the traditional placement of smartwatch faces. Yet, how the zones were used depended on the scenario. The \textcolor{BrickRed}{ulnar} zone was the least frequently reported for data representation, possibly due to the challenging arm posture it requires to access data. However, a few participants showed interest in utilizing this zone for decorative purposes and making fashion statements. For example, one participant described the decorative design on the \textcolor{BrickRed}{ulnar} zone) in her sketch (see~\autoref{fig:teaser}(b),\, (1): \emph{\quotes{For the spot (\textcolor{BrickRed}{ulnar} zone), I don't usually look at it; I want to make it decorative so that other people can notice and find it very beautiful.}} This finding suggests a categorization of the zones based on their accessibility: 1) the inner (or private) zones i.e., \textcolor{RoyalPurple}{radial} and \textcolor{ForestGreen}{volar} zones which are primarily viewed only by the wearer on demand (by rotating the forearm) and 2) the outer (or public) zones including, the \textcolor{CornflowerBlue}{dorsal} zone easily accessible for glanceable feedback to the wearer and potentially their surrounding, and the \textcolor{BrickRed}{ulnar} zone which is exposed to external viewers more than the wearers themselves. The different zones were very often outlined by separation lines (\autoref{fig:teaser}(b), (1), (3) and (6)\,), a chart (e.g., vertical progress bar like in \autoref{fig:teaser}(b), (7)\,, or simply white spaces (\autoref{fig:teaser}(b), (2)\,), which implies the importance of decorations or other graphical elements to create the overall aesthetics of the design.

\bpstart{Use space for detailed or supplementary information}
Surprisingly, people placed a similar amount of data items on the bands (6.5) compared to smartwatches (5) \cite{Islam:2020:Smartwatch-Survey,islam:2024:DesignSpace}. However, we found that participants used the available space on the prop to also include detailed views with more complex visualizations. For instance, the design in \autoref{fig:teaser}(b),  (2), shows a chart of fitness data on the smaller \textcolor{CornflowerBlue}{dorsal} zone over the last 10 hours. A more holistic view---over 24 hours---was drawn on the \textcolor{ForestGreen}{volar} zone.
 This finding aligns with the wearer's desire for more detailed visualization and going beyond the smartwatch display~\cite{Islam:2022:context-specific-smartwatch-vis}, as previously investigated in studies exploring the expansion of smartwatch output space and the integration of visually enhanced watchstraps~\cite{Klamka2020}. Especially the \textcolor{CornflowerBlue}{dorsal} and \textcolor{RoyalPurple}{radial} zones were used by participants for complementary and extended representations.
The \textcolor{CornflowerBlue}{dorsal} zone, for example, might display a summary of the activity (e.g., cycling speed of 10 km/h, distance traveled of 3 km, and cycling duration of 20 minutes), while the \textcolor{RoyalPurple}{radial} zone could show a map or route navigation as supplementary data. Combining multiple zones for data representation could offer added value to wearers in specific contexts. Future research may explore the optimal design of combined display zones beyond laboratory settings. 

\bpstart{Adjust visualizations based on postures}
While most participants preferred fixed data representation for contextual use, some emphasized the importance of responsiveness to context, suggesting that representation should dynamically adjust by rotating or resizing based on arm posture.  In particular, the ideas to change the rotation of representations seems useful to consider for future work. When participants can only briefly glance at a representation (e.g., cycling, driving) an up-right orientation with respect to their view may help to correctly and quickly read the data. 

\bpstart{Limitations and future work} While our study included several innovative ideas for the display of data representations on smart wristbands, the methodology we used is not without limitations. For example, we found that participants often suggested using text to represent data. In prior work, however, it has been shown that text was less effective than certain data representations on a smartwatch \cite{blascheck:hal-04018448}. In addition, while our findings highlight the potential for adaptable visualization designs to accommodate the wearers' arm movements and usage contexts, further research is necessary to which extent accommodation is beneficial. In addition, future studies need to explore the effectiveness of these design suggestions beyond laboratory settings. \revision{While we recognize these current limitations, our work does provide first starting points to understand and address the challenges of designing small data dashboards for smart wristbands that can guide and inspire future studies and development.}

\section {Conclusion}

In summary, our short paper points to several visualization design considerations for smart wristbands. While the technology is not yet commercially available, research prototypes are, and the visualization community can help to shape now how these wristbands will be used. In addition, research prototypes have already been built and can be used to evaluate open questions suggested here: how to detect postures and viewing angles, the usability and practical advantages of dynamically updating visualizations, perceptual challenges of using curved displays and viewing data across different zones, but also how to integrate data in designs that allow to make ``fashion statements.'' 

\begin{acks}
We thank the participants of our design workshop, Konstantin Klamka and Shaela Quader Ontu, for their feedback. The work was funded in part by ANR grant ANR-18-CE92-0059-01. Fairouz Grioui is funded by the Deutsche Forschungsgemeinschaft under Germany's Excellence Strategy -- EXC 2075 -- 390740016.
\end{acks}

\bibliographystyle{ACM-Reference-Format}
\bibliography{sample-base}



\end{document}